\documentclass[preprint,authoryear,12pt]{elsarticle}
\makeatletter
\def\ps@pprintTitle{%
 \let\@oddhead\@empty
 \let\@evenhead\@empty
 \def\@oddfoot{}%
 \let\@evenfoot\@oddfoot}
\makeatother

\usepackage{graphicx}
\usepackage{url}
\usepackage{amssymb}

\journal{New Astronomy}

\begin{document}

\begin{frontmatter}

\title{A Possible Converter to Denoise the Images of Exoplanet Candidates through Machine Learning Techniques}

\author[label1]{Pattana Chintarungruangchai}
\author[label1,label2]{Ing-Guey Jiang}
\ead{jiang@phys.nthu.edu.tw}
\author[label3]{Jun Hashimoto}
\author[label3]{Yu Komatsu}
\author[label4]{Mihoko Konishi}

\address[label1]{Department of Physics and Institute of Astronomy, National Tsing-Hua University, Hsin-Chu, Taiwan}
\address[label2]{Center for Informatics and Computation in Astronomy (CICA), National Tsing-Hua University, Hsin-Chu, Taiwan}
\address[label3]{National Astronomical Observatory of Japan (NAOJ), Mitaka, Tokyo, Japan}
\address[label4]{Faculty of Science and Technology, Oita University, Oita, Oita, Japan}

\begin{abstract}
The method of direct imaging has detected many exoplanets and made important contribution
to the field of planet formation. The standard method employs angular differential imaging (ADI) technique, 
and more ADI image frames could lead to the results with larger signal-to-noise-ratio (SNR).
However, it would need precious observational time from large telescopes,
which are always over-subscribed. We thus explore the possibility 
to generate a converter which can increase the SNR derived from a smaller number of ADI frames. 
The machine learning technique with two-dimension convolutional neural network (2D-CNN)
is tested here. Several 2D-CNN models are trained and their performances of
denoising are presented and compared. It is found that  
our proposed Modified five-layer Wide Inference Network 
with the Residual learning technique and Batch normalization 
(MWIN5-RB) can give the best result.
We conclude that this MWIN5-RB can be employed as a converter for future observational data.
\end{abstract}


\end{frontmatter}
\section{Introduction}
The subject of extra-solar planets (exoplanets) has been developed quickly due to
the continuous flow of new discoveries through several techniques.
It was particularly exciting to have directly imaged exoplanets for the first time in 2008 \citep{Kalas.Graham.2008}
as it gave signals directly from these exoplanets and thus confirmed their existence.
Since then, many research groups have spent considerable efforts to improve the techniques of 
high-contrast imaging in order to detect more exoplanets \citep{Tamura2009,EnyaAbe2010,Kuzuhara+2013, Dou+2015, DouRen2016}.
In addition, new high-contrast imaging instruments were developed for eight-meter class telescopes such as
the Gemini Planet Imager (GPI) \citep{Macintosh.Graham.2006} for Gemini South, 
the Subaru Coronagraphic Extreme Adaptive Optics (SCExAO) \citep{Jovanovic+2015} for Subaru Telescope,
and the Spectro-Polarimetic High contrast imager for Exoplanet Research (SPHERE) \citep{Beuzit.Vigan.2019} 
for Very Large Telescope (VLT). 
Moreover, a new camera was designed for SCExAO to further advance the performance of high contrast imaging \citep{Walter+2020}.
It is notable that these instruments often bring very interesting
related results \citep{Mayama+2006, Itoh+2008}.

To detect exoplanets through the method of direct imaging,
the high-contrast imaging employs the technique of angular differential imaging
(ADI) \citep{Marois.Lafreniere.2006} and produces many frames with different parallactic angles, i.e. ADI sequences.
Then, principle component analysis \citep{Soummer.Pueyo.2012,Amara.Quanz.2012} is used to
decompose the point spread function (PSF) of the central star from images taken with coronagraph in ADI sequences.
After that, the sparse signals of possible planets might be identified.
In order to reduce the noise among sparse signals,
\citet{GomezGonzalez.Absil.2016} considered one more component, i.e. the dense noise component,
and proposed an algorithm to extract exoplanet signals through a three-term decomposition.

Apparently, the more frames in ADI sequences,
the easier to get high signal-to-noise ratio (SNR) signals.
However, the number of frames of observations is restricted by
the resources of eight-meter class telescopes,
which are heavily used for many research fields.
Therefore, it is desirable to develop numerical procedures that 
could raise the SNR for a given number of frames in ADI sequences. 

Indeed, the machine-learning techniques have provided such possibilities, and
there are several methods which could reduce the image noises and have been used for non-astronomical purposes
\citep{Ronneberger.Fischer.2015,Mao.Shen.2016,Zhang.Zuo.2017}.
In fact, the machine learning has been employed in various fields in astronomy.
For example, it was used to search for exoplanet transits \citep{Pearson.Palafox.2018,Shallue.Vanderburg.2018,Chintarungruangchai.Jiang.2019,Yeh.Jiang.2021,Ofman+2022}. 
It was also used for the study of solar active regions \citep{Felipe.AsensioRamos.2019}, 
heliophysics \citep{Camporeale+2020},
remote sensing \citep{Lv+2019}, planetary science \citep{Lin+2018}, contact binaries \citep{Ding+2021}, 
stellar dynamics \citep{Liao+2022},
open clusters \citep{Gao2018}, young stellar objects \citep{Miettinen2018ApSS}, 
and galaxies \citep{Calleja2004, Schawinski.Zhang.2017, Cheng+2021}.  


Among many machine-learning techniques, the convolutional neural network (CNN) has been employed in many fields.
For example, \citet{Cheng+2021} employed the CNN to do the classification of galaxies,
\citet{Takahashi+2020} employed it to do the photometric classification of Hyper Suprime-Cam transients,
and recently \citet{Takahashi+2022} also used a deep CNN to do real/bogus classification for the Tomo-e Gozen transient survey.
In addition, \citet{SedaghatMahabal2018} performed image differencing with convolutional neural networks for real-time transient hunting.

The two-dimension convolutional neural network (2D-CNN) is often used for image processing, 
such as image recognition, object detection, semantic segmentation, and image conversion.
In general, the main task of image conversion is to convert one type of images into another type.
Thus, it could be used to convert noisy images into cleaner images, which is called a denoising process.

In this paper, we propose a new 2D-CNN technique to do the image denoising and increase the SNR for an exoplanet image. 
We would demonstrate that after some training by the data with more ADI-sequence frames of a given instrument, our 2D-CNN can increase SNR for the data with less ADI-sequence frames of the same facility.
For example, those observational data sets with 48 ADI-sequence frames can be taken 
to be the training data. For each data set, one low-quality image constructed from 20 ADI-sequence frames 
and another high-quality image constructed from 48 ADI-sequence frames can be made and fed into 
a 2D-CNN model (The choices of the above frame numbers would be explained in Section \ref{ssec:vltspheredataset}). 
This 2D-CNN model can learn the differences between low-quality and high-quality images.
After training by a huge number of low-quality and high-quality image pairs, this 2D-CNN model can then
be used to produce simulated high-quality images for those data sets only have 20 ADI-sequence frames.
Thus, this 2D-CNN model will be able to increase the probability of detecting exoplanets by the same telescope, 
which has limited observational time.

The 2D-CNN models are presented in Section \ref{sec:model}. The data preparation is mentioned in Section \ref{sec:data}.
The training process is described in Section \ref{sec:training}.
We then show the performance in Section \ref{sec:testing} 
and the demonstration in Section \ref{sec:demon}.
The conclusion remarks are provided in Section \ref{sec:conclu}.

\section{The 2D-CNN Models}
\label{sec:model}

There are many 2D-CNN models which are designed for 
image denoising.
The U-net has a U-shaped structure and employs both de-convolutional layers and convolutional layers \citep{Ronneberger.Fischer.2015} Its U-shape
produces additional direct connections between the outcomes of much earlier layers 
and the current layers. The connection is usually called ``concatenation''. 
The very deep Residual Encoder-Decoder Network (RED-Net)
is very similar to U-net but has a deeper structure. That is, it has many more layers \citep{Mao.Shen.2016}.

In addition, \citet{Zhang.Zuo.2017} presented the Denoising Convolutional Neural Network (DnCNN),
which included convolutional layers and batch normalization layers \citep{Ioffe.Szegedy.2015}.  
It does not have de-convolutional layers, and thus does not have a U-shape structure. 

Finally,  the five-layer Wide Inference Network with
the Residual learning technique and Batch normalization (WIN5-RB)
was proposed by \citet{Liu.Fang.2017}.
The model WIN5-RB uses a larger kernel size and a smaller number of layers
than other models. Thus, it has a wide and shallow structure.
WIN5-RB employs a residual learning technique, so
it learns from the difference between low-quality and high-quality images \citep{Kiku.Monno.2014}.
The structures of the above 2D-CNN models are summarized in Table \ref{tab:cnnmodel}.
As mentioned above, these models have different properties.
The numbers of layers of these four 2D-CNN models are different. 
They have been tuned and optimized as described in the sections of model architectures of those papers.

\begin{table*}
\centering
\caption{The Summary of 2D-CNN Models}
\begin{tabular}{lllll}
\hline
model & convolution & de-convolution & batchnorm & concatenation\\\hline
U-net & 18 & 5 & none & U-shape\\
RED-Net & 15 & 15 & none & U-shape\\
DnCNN & 17 & none & 15 & none\\
WIN5-RB & 5 & none & 5 & residual learning\\
\hline
\end{tabular}
\label{tab:cnnmodel}
\end{table*}

\begin{figure*}
\centering
\includegraphics[width=0.45\textwidth]{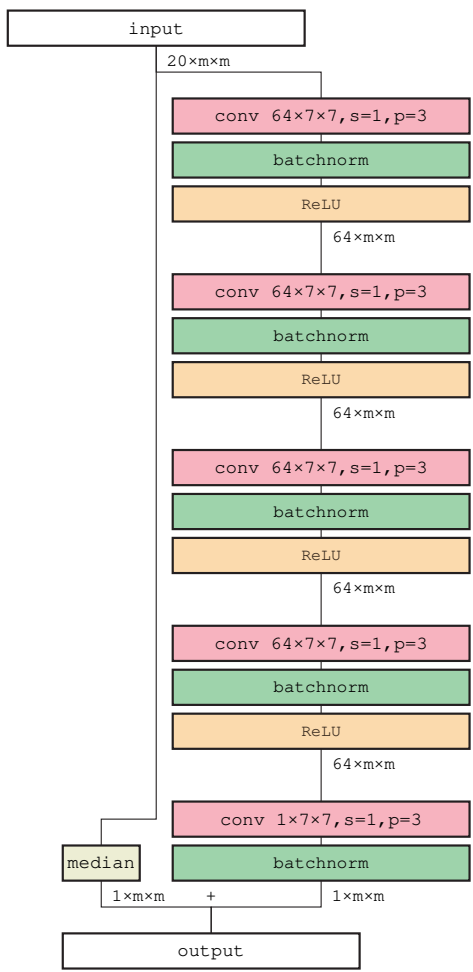}
\caption{The structure of MWIN5-RB.
The input is 20 images of an ADI sequence.
Those pink rectangular boxes (marked as conv) are convolutional layers.
The channel number$\times$kernel size is given right beside conv in these boxes.
In addition, $s$ is the striding size and $p$ is the padding size.
The width and height of images in the unit of pixel number is indicated by $m \times m$,
which is actually $256 \times 256$ in this paper.
The output from the final batchnorm layer is the "residual",
which would be added to the low quality image (i.e. the median of the input images) 
and lead to the output.}
\label{fig:mwin5rb}
\end{figure*}

\begin{figure}
\centering
\includegraphics[width=0.8\textwidth]{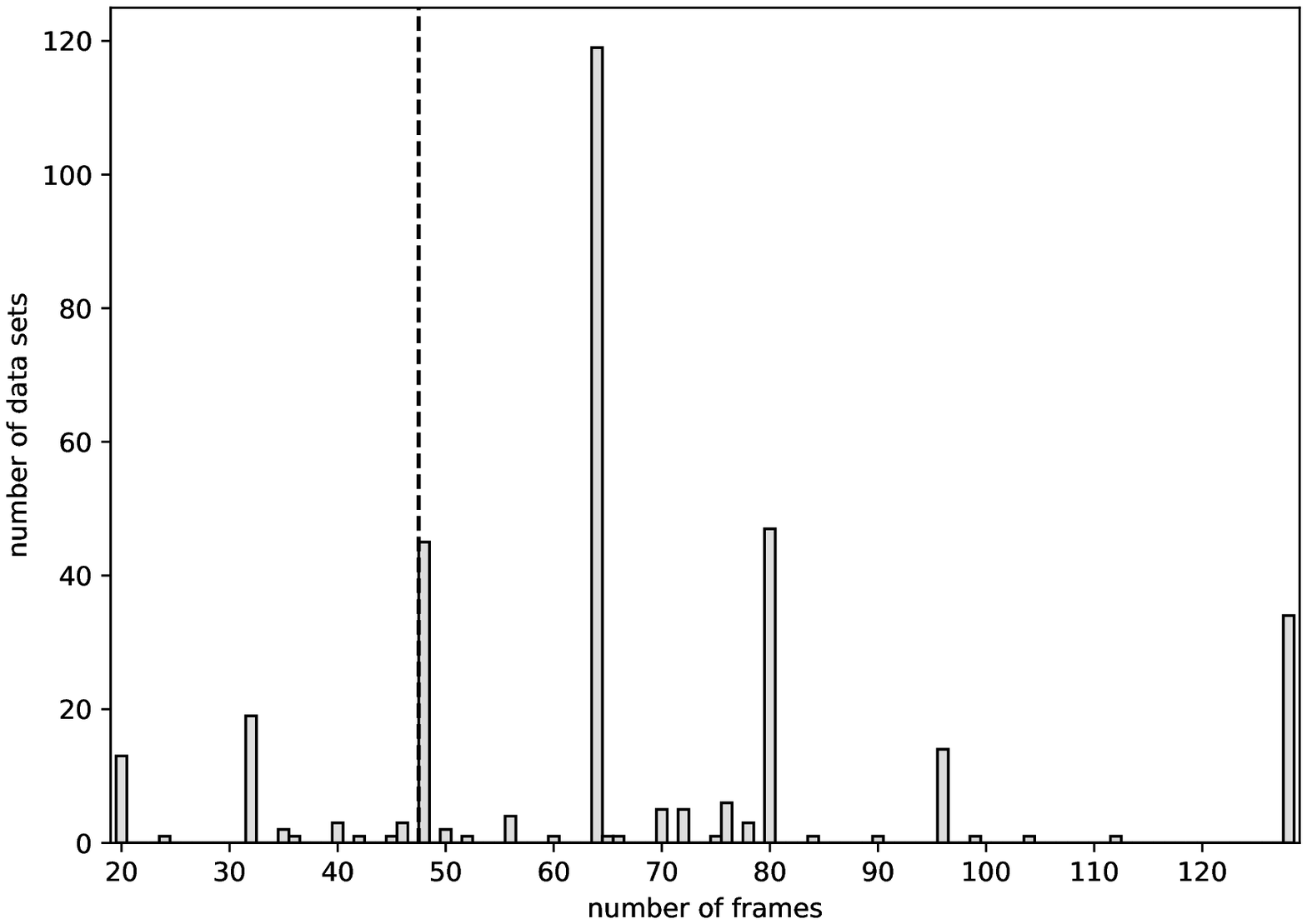}
\caption{The number of data sets as a function of frame numbers for the VLT SPHERE data employed in this work. 
The vertical dashed line corresponds to the frame number 48.} 
\label{fig:numfra}
\end{figure}

In order to choose a better 2D-CNN model for our purpose,  
we tried U-net, RED-Net, DnCNN and WIN5-RB and found that RED-Net 
took too much computer time due to its very deep structure.
After some consideration, we decided to create a new 2D-CNN model which is modified from 
WIN5-RB. It is thus called "Modified WIN5-RB" (MWIN5-RB), and  
its structure is shown in Figure \ref{fig:mwin5rb}.
With the same set of training data, the results of U-net, DnCNN, WIN5-RB,
and MWIN5-RB are compared, as can be seen in Section \ref{sec:testing}.
However, as the main goal of this paper is to propose MWIN5-RB and
present its performance, only its structure and equations,  
will be described in details.

As presented in Figure \ref{fig:mwin5rb},
the input of MWIN5-RB is 20 frames of ADI sequences after subtracting PSF and rotation 
(see details in Section \ref{ssec:combim}).
The input data will pass through 5 convolutional layers,
whose kernel size is $7 \times 7$, striding $s=1$, padding $p=3$,
each convolutional layer is followed by a batch normalization (batchnorm) layer
and a layer of rectified linear unit (ReLU) \citep{Hahnloser.Sarpeshkar.2000} as the activation function.
Note that there is no ReLU layer after the final batchnorm layer.
The details of these layers are described in Section \ref{ssec:operation}.

\section{The Data Preparation}
\label{sec:data}

To produce images which can be used as training and testing data, 
one synthetic planet is added on each observational image. 
In this paper, the images taken by
Spectro-Polarimetric High-contrast Exoplanet REsearch Instrument (SPHERE)
on Very Large Telescope (VLT) are employed.
Note that there are 10 known direct-imaging exoplanets among these data.
Because we will inject synthetic planets into images to produce our training data, 
these data with detected exoplanets will not be used.

\subsection{VLT SPHERE Data}
\label{ssec:vltspheredataset}

The observational data of VLT SPHERE can be downloaded from the European Southern Observatory (ESO) archive website\footnote{ http://archive.eso.org/eso/eso\_archive\_main.html}.
The downloaded data are raw data, which also include data for calibration (flat, dark, sky).
We do pre-processing for SPHERE data by the python module VLT/SPHERE \citep{Vigan.2020}.
This pre-processing includes dividing flat, subtracting background, correcting bad pixel, and re-centering.

SPHERE has many observation modes,
and here we make use of the data which was taken in dual-band imaging mode \citep{Vigan.Moutou.2010}
with coronagraph by Infra-Red Dual-beam Imager and Spectrograph (IRDIS) \citep{Dohlen.Langlois.2008}.
We choose those SPHERE-IRDIS data which was observed from 2015 to 2018 
with broadband H filter (BBH) and dual band H2-H3 filter (DBH23).
The exposure times of these data are 16, 32, and 64 seconds.
The number of image frames for one data set varies from a few to more than one hundred, i.e.
from 2 to 128.
After excluding those data sets that have less than 20 frames, 
those with known exoplanets, and those failed sets during data reduction,
there are 395 data sets. In order to select from these data sets for  
the further training and testing, the histogram of frame number of these 395 data sets
is presented in Figure \ref{fig:numfra}. 
The vertical dashed line in Figure \ref{fig:numfra} indicates where the frame number is 48.
From the figure, we can see that there are many data sets that have 48 frames.
We choose the data sets that have 48 or more frames to be the training and testing data.
There are 351 such data sets. In addition, the reason we choose 20 to be the frame number of low quality image
is that the SNR difference between low and high quality images could be large enough for our study
(please see the results in Figure \ref{fig:snr_frame}). 

There are two IRDIS images taken simultaneously for each observation runs. 
Therefore, we actually have 702 data sets.
Because the training is the most important part of a machine-learning technique, it is common to use 
about 80 percents of available samples for the training and 20 percents for the testing.
We split the 702 data sets into 2 parts: 602 samples for the training and 100 samples for the performance testing.

\subsection{Injecting Planet}
\label{ssec:inject}

Because a huge number of training samples is needed for a machine learning technique,
we inject one synthetic planet into each image randomly. 
Thus, the position and brightness of the planet in each image is known 
for the training and testing data.

For the VLT SPHERE data, in addition to the ADI sequences taken with coronagraph,
there is an image taken by moving the telescope's pointing far from the original view center
in order to have the true PSF of the central star without the coronagraph effect.
The normalized PSF derived from the host-star images without coronagraph 
is employed to be the PSF of a synthetic planet.
Noted that this PSF is different from the PSF that we will obtain from ADI sequences 
which were taken using coronagraph.

The size of calibrated image is 800 pixel $\times$ 800 pixel. 
Considering the star to be at the image center and using pixel size as the length unit, the planet is injected randomly in an annulus area with radial coordinate $r_p \in [50, 200]$ following a uniform distribution. 
After the planet's radial coordinate $r_p$ is given, the planet's PSF is re-scaled by multiplying a factor.  
This factor is randomly set to be within the range $[2,4]$ times of 
the standard deviation of image's flux values in the area with 
radial coordinate $r \in [r_p - FWHM, r_p + FWHM]$, where 
FWHM is the full width at half maximum of the planet's PSF.
The synthetic planet in each image frame is placed at different angle based on the parallactic angle of that frame,
so that the planet position would be the same after rotating back by each parallactic angle.
The position and brightness of synthetic planets are re-generated randomly for each training epoch.

\subsection{Combining Images}
\label{ssec:combim}
In order to make a pair of low and high quality images for the purpose of training and testing,
20 frames are combined to form a low quality image, and
48 frames are combined to produce a high quality image.
To proceed the combination, we need to do PSF subtraction and the rotation of parallactic angle.

To obtain PSF and subtract it from each image, we use principle component analysis (PCA),
which is also called Karhunen-Loève image projection (KLIP),
as that used by \citet{Soummer.Pueyo.2012} and \citet{Amara.Quanz.2012}.
After PSF subtraction,
each image is rotated depending on the parallactic angle of each image.
Then, by finding the median, 20 images are combined to be the low quality image, 
and 48 images are combined to be the high quality image.

\begin{figure*}
\centering
\includegraphics[width=1\textwidth]{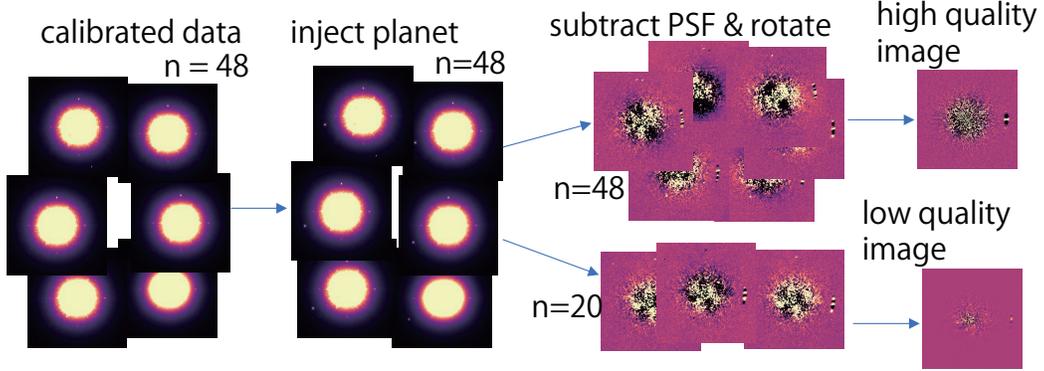}
\caption{The process of data preparation.
Each calibrated data is injected one planet.
Then, 20 images are used to calculate the PSF of low quality image
and 40 images are used to calculate the PSF of high quality image.
After subtracting PSF and rotating images by parallactic angle,
the corresponding median for low and high quality images are obtained.
}
\label{fig:prepare}
\end{figure*}

All images are resized from $800 \times 800$ pixels into $400 \times 400$ pixel by interpolation,
and then we crop only $256 \times 256$ pixels at the center.
The flow chart of data preparation is shown in Figure \ref{fig:prepare}.

\subsection{Signal-to-Noise Ratio}
\label{ssec:snr}

We need to determine the SNR of both low quality and high quality images.   
As in \citet{Mawet.Milli.2014} and \citet{GomezGonzalez.Absil.2016},
SNR is calculated by below equation
\begin{eqnarray}
\label{equ:snr}
\mathrm{SNR} = \frac{\bar{x}_1-\bar{x}_2}{s_2\sqrt{1-\frac{1}{n_2}}},
\end{eqnarray}
where $\bar{x}_1$ is the mean intensity in the circle with radius $\mathbf{FWHM}/2$ around the considered point.
In addition, $n_2$ is the number of all circles at the same radius $(n_2=round(2\pi / \theta)-1)$,
where $\theta = 2\arcsin{\left(\frac{\mathrm{FWHM/2}}{r}\right)}$,
and $r$ is the radial coordinate of that considered point,
$\bar{x}_2$ and $s_2$ are the mean intensity and the empirical standard deviation
computed over these $n_2$ circles.

After injecting a planet, we calculate the SNR of the injected planet in the image.
The SNR of high quality images should be higher than the SNR of low quality images.
However, the SNR of low quality image is higher than the SNR of high quality image for some injected planets occasionally.
These images are excluded as they are not appropriate to be our training or testing data.
Furthermore, we also exclude the data that the SNR of low quality image is lower than 2 or higher than 15,
and that the SNR of high quality image is lower than 5 or higher than 100.
In this paper, the SNR of an image means the SNR value of the planet in this image,
and the SNR map of an image means the values of SNR 
for pixels in an area with radial coordinate $r \in [12, 100]$ in this image.

\begin{figure}
\centering
\includegraphics[width=0.8\textwidth]{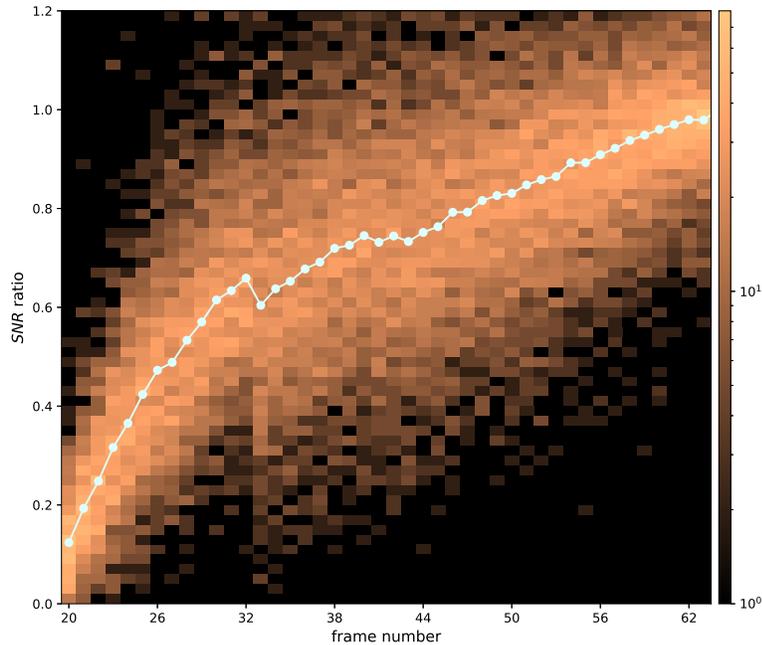}
\caption{The SNR (normalized to the SNR of 64 frames) as a function of frame number.
The color indicates the number of data sets and the white curve with points show the median.
}
\label{fig:snr_frame}
\end{figure}

The value of SNR is higher when the image is combined from more frames of a data set of ADI sequence.
Figure \ref{fig:snr_frame} shows this effect.
To make this figure, here we inject planet and combine images as explained in Section \ref{ssec:combim},
but with different number of combined frames, which are in the range from 20 to 64 frames.
The $x$-axis is the number of combined frames and
the $y$-axis is the SNR value normalized to the SNR of the image combined from 64 frames.
The color indicates the number of data sets of a particular point on the $x-y$ plane.
The white curve and the points show the median of the SNR of all training data sets for a given number of combined frames.
We can see that the SNR increases with the frame number.

\section{The Training}
\label{sec:training}

The Pytorch module \citep{Paszke.Gross.2019}
written in python language was employed
to build the CNN models in this work.
These CNN models will be trained by the prepared training data.
During the training, the fluxes of input images are taken by a CNN model and a predicted 
high quality image is obtained. Through the comparison between this predicted high quality image 
and the original high quality image,  the parameters of the CNN model are tuned and optimized. 
One training epoch means that all training data are used to train a CNN model for one time.
The whole process, including the planet injections of data preparation, would be repeated
and thus another training epoch can be done. 

\subsection{The Input}
\label{ssec:input}

There are 602 data sets for each training epoch.
We train the model by separating the data into groups, i.e. batches.
The number of data sets in each batch is 32 (except the final batch).
Let $x_{0*}^{(t,j,u,v)}$ be the value of each point on one image frame, 
where $t$ is the index of each data set in one batch, $j$ is the index of each ADI frame in one data set,
$u$ and $v$ are the row and column index of an image pixel.
In addition, the values of each point of the high quality images combined from 48 frames of an ADI sequence are also needed
as they will be used during the optimization stage of the training process. 
We use $y_{*}^{(t,u,v)}$ to denote these values, where $t,u,v$ are still the indexes of data set, row, and column, respectively.
Because the minimum and maximum values on
each image frame are different, a normalization process is necessary. 
Before entering any layer of a 2D-CNN Model,
a shifting and scaling for the above values are performed through 
\begin{equation}
\label{equ:scaX}
x_0^{(t,j,u,v)} = \frac{x_{0*}^{(t,j,u,v)} - \mathbf{med}_{j,u,v}(x_{0*}^{(t,j,u,v)})}
{6\ \mathbf{mad}_{j,u,v}(x_{0*}^{(t,j,u,v)})}, 
\end{equation}

\begin{equation}
\label{equ:scaY}
y^{(t,u,v)} = \frac{y_{*}^{(t,u,v)} - \mathbf{med}_{u,v}(y_{*}^{(t,u,v)})}
{6\ \mathbf{mad}_{j,u,v}(x_{0*}^{(t,j,u,v)})},
\end{equation}
where $\mathbf{med}_{j,u,v}( )$ is the median of values over all $j, u, v$,
$\mathbf{med}_{u,v}( )$ is the median of values over all $u, v$, and
\begin{eqnarray}
\label{equ:mad}
& &\mathbf{mad}_{j,u,v}(x_{0*}^{(t,j,u,v)}) \nonumber \\
&=&\mathbf{med}_{j,u,v}(|x_{0*}^{(t,j,u,v)}-\mathbf{med}_{j,u,v}(x_{0*}^{(t,j,u,v)})|).
\end{eqnarray}

Note that U-net, DnCNN, WIN5-RB, use the median of 20 images as input, 
but MWIN5-RB directly use 20 images as input.  

\subsection{The Operations of Layers}
\label{ssec:operation}

Although all four 2D-CNN models, i.e. U-net, DnCNN, WIN5-RB, and MWIN5-RB, are trained and 
all their performance will be presented, here we mainly describe the calculation details of the MWIN5-RB model.
This is because MWIN5-RB is the new one we propose here and its full self-consistent description is necessary.
The structures of other 2D-CNN models can be obtained from the corresponding reference papers.

Following the structure of MWIN5-RB in Figure \ref{fig:mwin5rb},
the input data will go through convolutional layers,  batchnorm layers, and ReLU layers.
Set $h_n^{(t,j,u,v)} = \mathbf{conv}(x_n^{(t,j,u,v)})$,
where $\mathbf{conv}$ denote the operation of a convolutional layer.
Let $\mathbf{batchnorm}$ denote the operation of 
a batchnorm layer, $\mathbf{ReLU}$ be the operation of a ReLU layer.
We have
\begin{eqnarray}
\label{equ:dncnn}
x_{n+1}^{(t,j,u,v)} = \mathbf{ReLU}(\mathbf{batchnorm}(h_n^{(t,j,u,v)})),
\end{eqnarray}
where $n=0,1,2,3$, and $t,u,v$ are as defined previously.
Note that $j$ is the index of each ADI frame for $x_0$ but for all layers $j$ is channel index of output of that layer.
The above recursive calculation gives an outcome $x_4^{(t,j,u,v)}$. 
With $h_4^{(t,j,u,v)} = \mathbf{conv}(x_4^{(t,j,u,v)})$, the next outcome is obtained as
\begin{eqnarray}
x_5^{(t,u,v)} = \mathbf{batchnorm}(h_4^{(t,j,u,v)}). 
\end{eqnarray}

The equation of the $n^{th}$ convolutional layer is
\begin{eqnarray}
\label{equ:convol}
& &h_n^{(t,j,u,v)}   \nonumber  \\   
&=&\sum_{i=1}^{c_{n-1}}\sum_{l=1}^{k_n}\sum_{m=1}^{k_n}w_n^{(
i,j,l,m)}x_n^{(t,i,u+l-p_{n}-1,v+m-p_{n}-1)}+b_n^{(j)} \nonumber \\
& &
\end{eqnarray}
when
\begin{eqnarray}
\label{equ:conwhen}
& &p_{n} < u \le s_{1,n}-p_{n}, \nonumber \\
& &p_{n} < v \le s_{2,n}-p_{n}
\end{eqnarray}
and $h_n^{(t,j,u,v)} = 0$ for the rest,
where $w_n^{(i,j,l,m)}$ is the weight parameter of point $(l,m)$ of kernel for input channel $i$ and output channel $j$,
$b_n^{(j)}$ is the bias parameter of output channel $j$,
$c_n$ is the channel number, $k_n$ is the kernel size, and $p_n$ is the padding size.
For U-net and DnCNN, $k_n = 3$ and $p_n=1$. For WIN5-RB and MWIN5-RB, $k_n = 7$ and $p_n=3$.
In addition, $s_{1,n}$ is the row pixel number and $s_{2,n}$ is the column pixel number of an image.
Note that $w_n^{(i,j,l,m)}$ and $b_n^{(j)}$ are trainable parameters. 
They are trained and changed during the training process by stochastic gradient descent (SGD).

The batch normalization layer \citep{Ioffe.Szegedy.2015} is commonly used in neural networks.
It normalizes the output of the previous layer by re-centering and re-scaling,
and thus improves the training speed.
The equation of the batch normalization layer is
\begin{eqnarray}
\label{equ:batchnorm1}
r_n^{(t,j,u,v)} &=& \mathbf{batchnorm}(h_n^{(t,j,u,v)}) \nonumber \\ 
&=& \gamma_n\left(\frac{h_n^{(t,j,u,v)} - \mu_{n,B}^{(j)}}{\sqrt{(\sigma_{n,B}^{(j)})^2+\epsilon}} \right) + \beta_n,
\end{eqnarray}
where $\epsilon = 10^{-5}$ is a small value added for numerical stability.
$\gamma$ and $\beta$ are parameters that have to be trained and usually changed by SGD during the training.
$\mu_{n,B}^{(j)} = \overline{h_n^{(t,j,u,v)}}$ is the mean of $h_n^{(t,j,u,v)}$ in channel $j$ for all $t,u,v$ in the batch,
and $(\sigma_{n,B}^{(j)})^2$ is the variance, i.e.
\begin{eqnarray}
\label{equ:batchnorm2}
(\sigma_{n,B}^{(j)})^2 = \overline{\left(h_n^{(t,j,u,v)} - \mu_{n,B}^{(j)}\right)^2}.
\end{eqnarray}

Finally, the equation of the ReLU layer \citep{Hahnloser.Sarpeshkar.2000} is 
\begin{eqnarray}
\label{equ:relu}
x_{n+1}^{(t,j,u,v)} = \mathbf{ReLU}(r_n^{(t,j,u,v)}) = \max(0,r_n^{(t,j,u,v)}).
\end{eqnarray}

\subsection{The Residual}
The residual implementation is used only in WIN5-RB and MWIN5-RB, not in U-net and DnCNN.
The output from the last convolutional layer of WIN5-RB and MWIN5-RB is called "residual",
which can be interpreted as the difference of values between low quality image and predicted high quality images.
The residual will be added to low quality images and the predicted high quality images 
can be obtained.

Note that, in MWIN5-RB, the input is 20 images (i.e. 20 frames),
the low quality sample is the median-combined image of these 20 frames,
while the high quality image which is used during the optimization 
is the median-combined image of 48 frames.

The output for our model is
\begin{eqnarray}
\label{equ:residual}
z^{(t,u,v)} = \mathbf{med}_j(x_0)^{(t,j,u,v)} + x_5^{(t,u,v)}
\end{eqnarray}
where $\mathbf{med}_j(x_0^{(t,j,u,v)})$ is the median over all channel $j$ of the input $x_0^{(t,j,u,v)}$,
$x_5^{(t,u,v)}$ is the output of the last layer.
Note that the input is 20 images before combining, used as 20 channels,
so it is 4 dimensional data that include an index of channel $j$.
Whereas the output from last layer $x_5^{(t,u,v)}$ (residual) and the high quality image $y^{(t,u,v)}$,
and also the last output (i.e. residual + low quality image) $z^{(t,u,v)}$,
are 3 dimensional data without index of the channel $j$.

\subsection{The Optimization}
\label{ssec:optim}

The input images pass through all layers of CNN models and the output is obtained.
Then, we calculate the loss as mean square error (MSE) between the output $z^{(t,u,v)}$ 
and the original high quality image $y^{(t,u,v)}$.
The loss is weighted by a Gaussian function centered at the planet position as
\begin{eqnarray}
\label{equ:mse}
D^{(t,u,v)}&=&(y^{(t,u,v)}-z^{(t,u,v)})^2\exp{\left(-\frac{1}{2}\left(\frac{R^{(t,u,v)}}{2\mathbf{FWHM}^{(t)}_{\mathbf{PSF}}}\right)^2\right)}, \nonumber \\
\mathbf{MSE}&=&\mathbf{mean}(D^{(t,u,v)}),
\end{eqnarray}
where $R^{(t,u,v)}$ is the radial coordinate of the planet position, $\mathbf{FWHM}^{(t)}_{\mathbf{PSF}}$ is the FWHM of the PSF.
The partial derivative of the MSE with respect to a given parameter $P_{\tau}$ is
\begin{equation}\label{equ:partialdiff}
g_{\tau} = \frac{\partial \mathbf{MSE}}{\partial P_{\tau}},
\end{equation}
where $\tau$ is the batch index, $P_{\tau}$ is any parameter to be optimized, including $w_n^{(i,j,u,v)}$ and $b_n^{(j)}$ in each convolutional layer and $\beta_n$ and $\gamma_n$ in each batch normalization layer.
With the partial derivative $g_{\tau}$, each parameter is optimized through Adam optimization algorithm \citep{Kingma.Ba.2014}.
During the training process, each parameter is updated as
\begin{equation} \label{equ:adam}
P_{\tau} = P_{\tau-1}-\eta \frac{m_\tau}{1-\beta_1^\tau}
\left(\sqrt{\frac{v_\tau}{1-\beta_2^\tau}}+\epsilon\right)^{-1},
\end{equation}
where $\beta_1$, $\beta_2$ and $\epsilon$ are Adam's parameters, and
\begin{eqnarray}\label{equ:adammv}
m_\tau &=& m_{\tau-1} + (1-\beta_1)g_{\tau-1}, \nonumber  \\
v_\tau &=& v_{\tau-1} + (1-\beta_2)g_{\tau-1}^2,
\end{eqnarray}
where $m_0=v_0=0$.
We use the above Adam optimization with
$\eta=0.001$, $\beta_1=0.9$, $\beta_2=0.999$, $\epsilon=10^{-8}$.

\section{The Resulting Performance}

\label{sec:testing}
After all four models are trained, the testing data would be employed to test their performance.
The testing data are read into the trained CNN models and the predicted high quality images
would be obtained as the outcome of these models. 

The calculations during the training phase and the testing phase are the same for convolution layers, ReLU layers, 
the residual determination, but are different for batch normalization layers.
In testing phase, the calculation in Eq. \ref{equ:batchnorm1} of batch normalization layers (for DnCNN, WIN5-RB, and MWIN5-RB) 
is replaced by
\begin{eqnarray}
\label{equ:batchnorm3}
r_n^{(t,j,u,v)} &=& \mathbf{batchnorm}(h_n^{(t,j,u,v)}) \nonumber \\
		&=& \gamma_n\left(\frac{h_n^{(t,j,u,v)} - \mu_{n,R}^{(j)}}{\sqrt{(\sigma_{n,R}^{(j)})^2+\epsilon}} \right) + \beta_n.
\end{eqnarray}
Here 
$\mu_{n,R}^{(j)}$ and $(\sigma_{n,R}^{(j)})^2$ are the final values of the below 
iterations calculated during the training phase:
\begin{eqnarray}
\label{equ:batchnorm4}
\mu_{n,R,\tau}^{(j)} &=& \nu\mu_{n,R,\tau-1}^{(j)} + (1-\nu)\mu_{n,B,\tau}^{(j)} \nonumber \\
(\sigma_{n,R,\tau}^{(j)})^2 &=& \nu(\sigma_{n,R,\tau-1}^{(j)})^2 + (1-\nu)(\sigma_{n,B,\tau}^{(j)})^2,
\end{eqnarray}
where $\nu$ is the momentum, which is set to be $\nu=0.9$ here. The index $\tau$ is the batch identity which would always keep 
increasing for all epochs during the training phase.
Note that $\tau$ is an integer starting from $\tau=1$, and initially  
$\mu_{n,R,0}^{(j)} \equiv \mu_{n,B,1}^{(j)}$ and $(\sigma_{n,R,0}^{(j)})^2 \equiv (\sigma_{n,B,1}^{(j)})^2$.

When the training stops, 
$\mu_{n,R}^{(j)}$ and $(\sigma_{n,R}^{(j)})^2$ are then obtained as the final values of 
$\mu_{n,R,\tau}^{(j)}$ and $(\sigma_{n,R,\tau}^{(j)})^2$. 
Therefore, both $\mu_{n,R}^{(j)}$ and $(\sigma_{n,R}^{(j)})^2$ are constants during the testing phase.
In addition, since the training has finished and the trained CNN models are obtained, there is no optimization here.  

\begin{figure}
\centering
\includegraphics[width=0.8\textwidth]{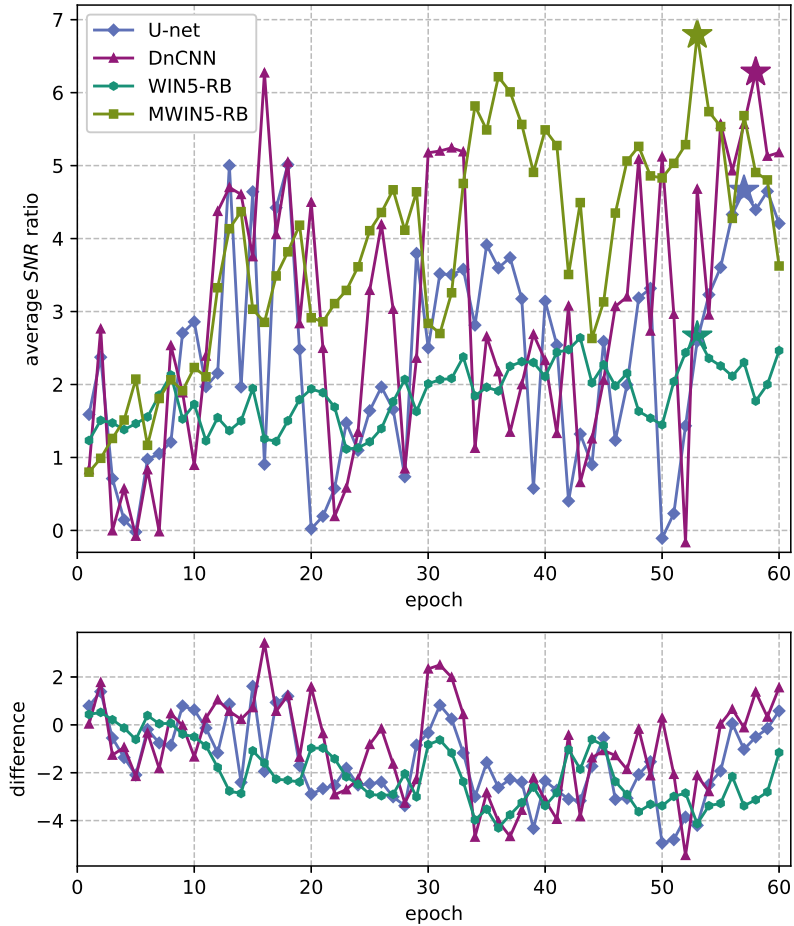}
\caption{The top panel is the average SNR ratio of the predicted high quality image over the low quality image (SNR$_{predict}$/SNR$_{low}$)
of the testing data after each training epoch.
The maximum values of each model are indicated by the starred points.
The bottom panel shows the difference that a particular model's (indicated by the 
same color in the top panel) average SNR ratio is 
subtracted by the MWIN5-RB's.
}
\label{fig:snr_epoch}
\end{figure}

In order to quantify the performance of CNN models, the SNR of the predicted high quality image is calculated.
Correspondingly, the SNR of the median of 20 input images is also calculated.  
The SNR ratio defined as the predicted one over the input one is then determined. 
The larger SNR ratio indicates the better denoising performance of a CNN model. 
CNN models could become better denoising converters if more training epochs are done. 
To understand when the training shall stop, 
The top panel of Figure \ref{fig:snr_epoch} shows the average SNR ratio of the results from 100 testing data sets
as a function of the number of training epochs for the considered four CNN models. 
It is found that the performance is saturated when there are about 60 training epochs.
The peaks, as indicated by the starred symbols, 
are 57 epochs for U-net, 58 epochs for DnCNN, and 53 epoch for WIN5-RB and MWIN5-RB.
Thus, the trained CNN models with the above number of training epochs would be used 
as standard models for all the results hereafter.
It is shown that the peak of MWIN5-RB is higher than 6.5. 
This value is actually close to the SNR ratio of the original high quality image over the input low quality image.
In order to compare U-net, DnCNN, WIN5-RB with our MWIN5-RB, the bottom panel of Figure\ref{fig:snr_epoch} presents the difference that a particular model's (indicated by the 
same color in the top panel) average SNR ratio is 
subtracted by the MWIN5-RB's. Note that the results of early epochs are not important as all models
are still under the early training stage.  It is the peak while the performance is saturated matters.


\begin{figure*}
\centering
\includegraphics[width=1\textwidth]{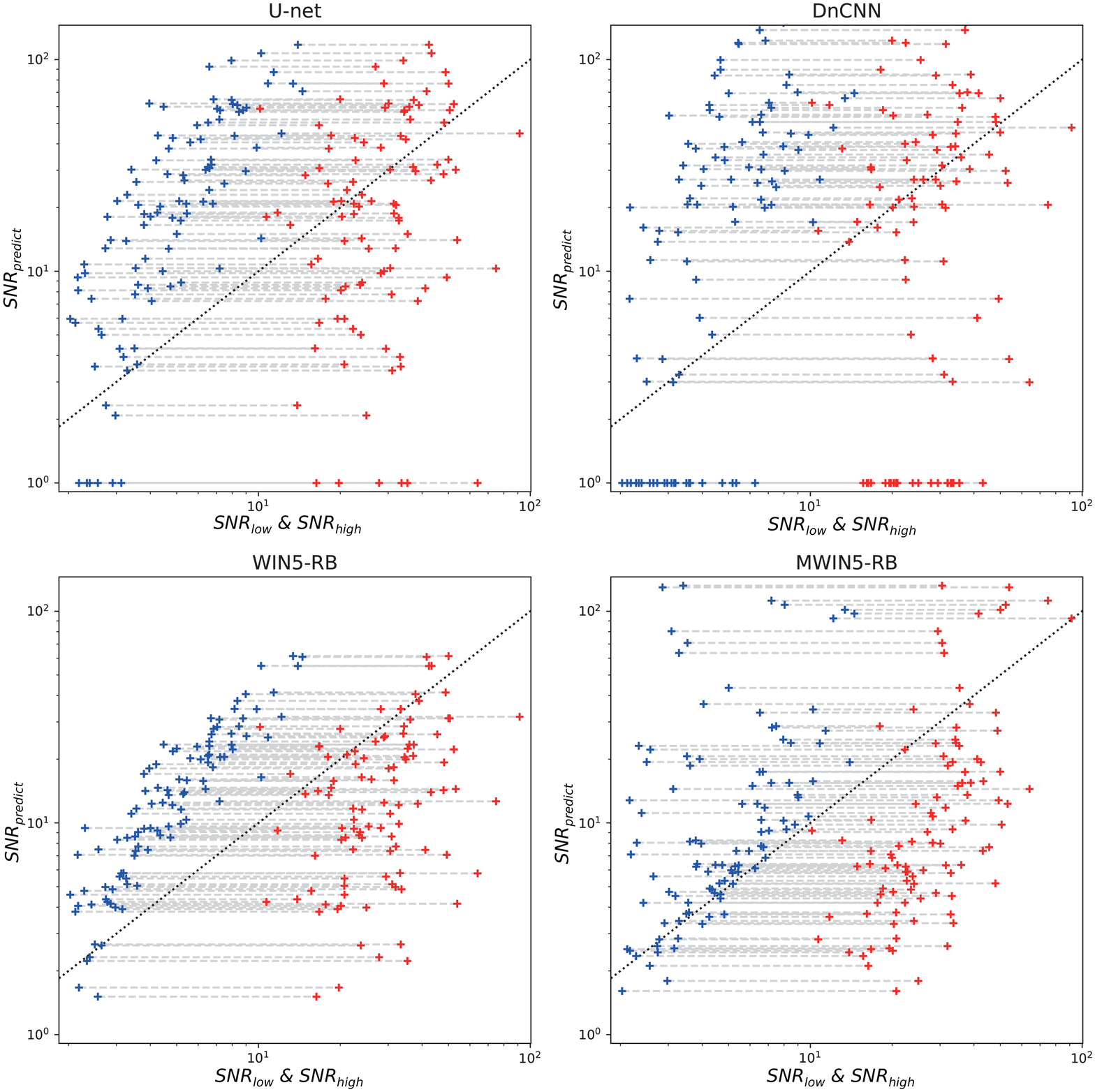}
\caption{The SNR of low quality image (blue) and original high quality image (red) versus 
the SNR of predicted high quality image for all testing data.
The dashed line is the border where SNR$_{predict}$=SNR$_{low}$ (or SNR$_{high}$)}
\label{fig:snr_scatter}
\end{figure*}

In addition, a scatter plot of SNR for all 100 testing data sets is presented in Figure \ref{fig:snr_scatter}.
In this plot, $x$-axis is the SNR of input low quality image ($SNR_{low}$, i.e. blue points) 
and the SNR of original high quality image ($SNR_{high}$, i.e. red points), 
$y$-axis is the SNR of predicted high quality image ($SNR_{predict}$).
Note that the scatter plot is drawn in log scale,
and all points with $SNR_{predict} < 1$ are plotted at $SNR_{predict}=1$.
The tilted dotted line is $y=x$, which indicates the positions where $SNR_{low}$ or $SNR_{high}$ equals to $SNR_{predict}$.
The horizontal lines simply connect blue and red points of the same data sets.   
The positions of blue or red points relative to the tilted line indicate 
whether the predicted high quality images has larger SNR than the input low quality image or
the original high quality image. 


For the MWIN5-RB model, out of 100 testing data sets,  
86 data sets have the results $SNR_{predict} > SNR_{low}$ and  
14 data sets have the results $SNR_{predict} < SNR_{low}$.
Among those 86 successful cases, there are 13 data sets that the predicted 
high quality images even have larger SNR than the original high quality images,
i.e. $SNR_{predict} > SNR_{high}$.
Therefore, the under-predicted rate is 14/100, i.e. 14 percents, 
and the over-predicted rate is 13/100, i.e. 13 percents.

\begin{figure}
\centering
\includegraphics[width=0.75\textwidth]{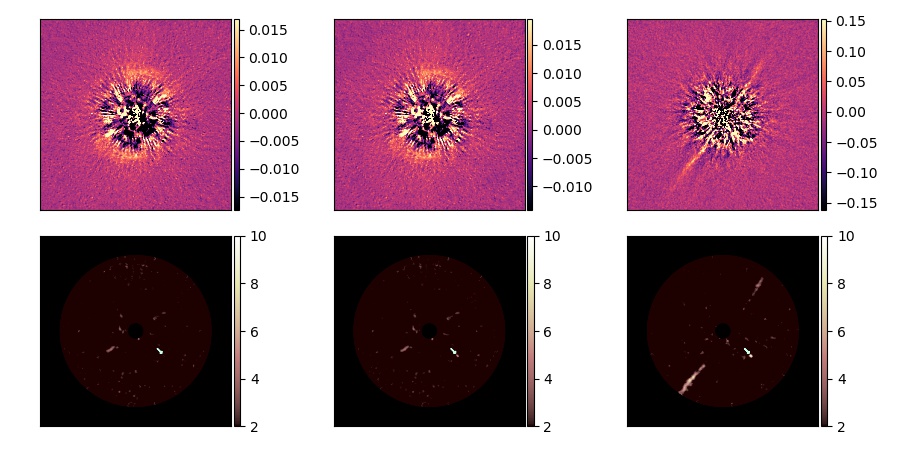}
\caption{The result of a chosen testing data in MWIN5-RB. 
Upper rows are images and lower rows are SNR maps.
Left panels are for the low quality image,
central panels are for the predicted high quality image,
and right panels are for the original high quality image.
The arrow indicates the planet's position.
From left to right, the SNRs of the upper panels
are 3.787, 8.271, 13.100, respectively.
}
\label{fig:risa1}
\end{figure}

\begin{figure}
\centering
\includegraphics[width=0.75\textwidth]{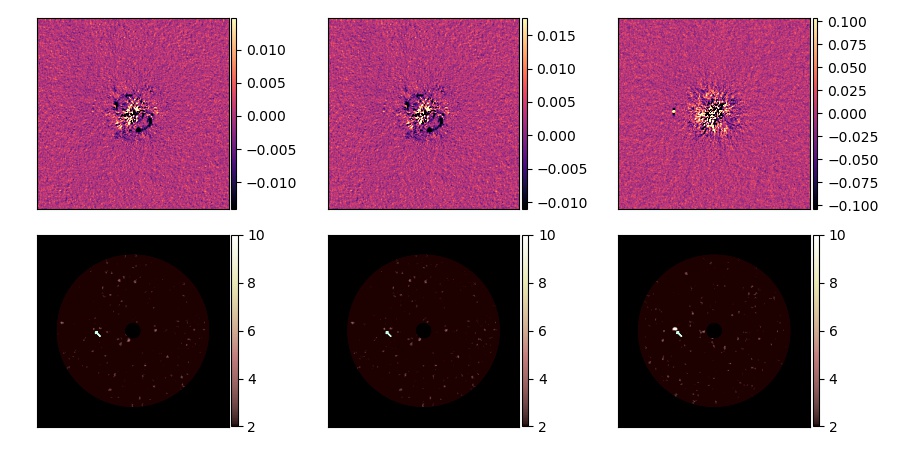}
\caption{The same as Figure \ref{fig:risa1}, but for a case whose SNR does not really increase.
From left to right, the SNRs of the upper panels
are 4.443, 4.669, 28.993, respectively.
}
\label{fig:risa2}
\end{figure}

\begin{figure}
\centering
\includegraphics[width=0.75\textwidth]{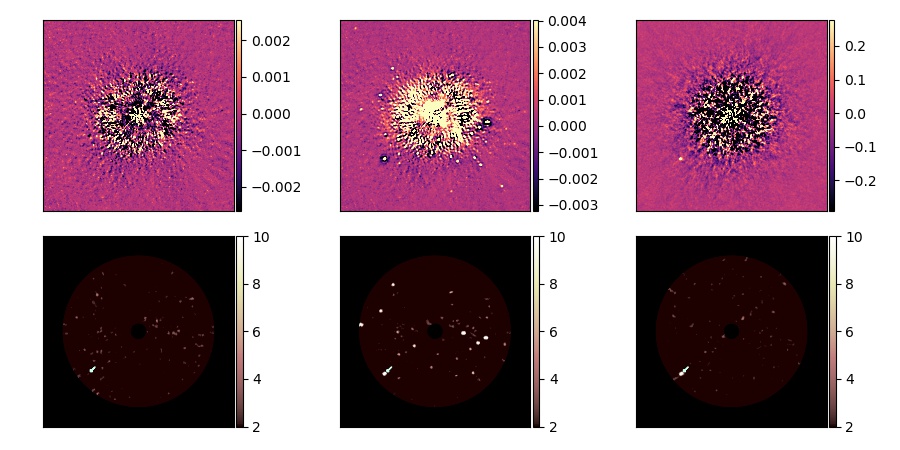}
\caption{The same as Figure \ref{fig:risa1}, but for a case whose SNR is over-predicted.
From left to right, the SNRs of the upper panels
are 3.532, 70.870, 30.551, respectively.
}
\label{fig:risa3}
\end{figure}

Figure \ref{fig:risa1} shows the images and SNR maps of one chosen testing data in MWIN5-RB.
The images are on the upper row and the corresponding SNR maps are on the lower row.
The left panels are for the low quality image,
the central panels are for the predicted high quality image,
and the right panels are for the original high quality image.
The cyan arrow indicates the planet's position.
This case represents the successful majority obtained by MWIN5-RB that 
$SNR_{predict} > SNR_{low}$, and $SNR_{predict}$
is close to but smaller than $SNR_{high}$.

In addition, Figure \ref{fig:risa2} is for the case that SNR does not really increase, 
and Figure \ref{fig:risa3} is for the over-predicted case that 
$SNR_{predict} > SNR_{high}$. It shows that 
many points of false planets can appear in the SNR map for this over-predicted case.

\begin{figure}
\centering
\includegraphics[width=0.75\textwidth]{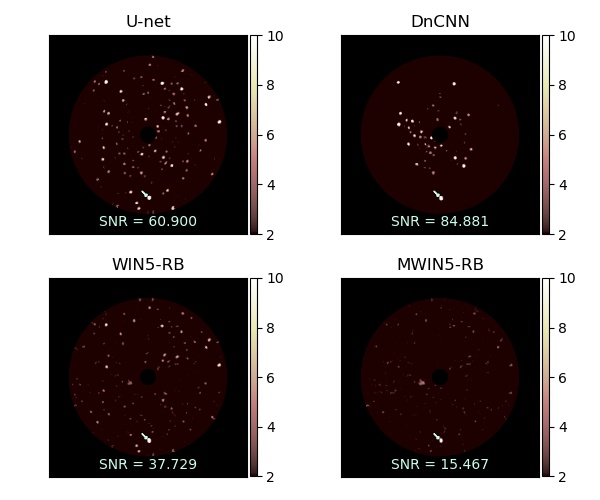}
\caption{The SNR maps of predicted high quality images for a given testing data obtained by four CNN models. 
}
\label{fig:snrmap_compare}
\end{figure}

Moreover, Figure \ref{fig:snrmap_compare} provided SNR maps of predicted high quality images for one testing data set 
obtained by all considered CNN models for the purpose of comparison.  
It was found that the false planet problems are serious for DnCNN, U-net, and WIN5-RB models.
Therefore, we propose that our MWIN5-RB model is the best here and would do further demonstration 
for it in the next section.

\section{The Demonstration}
\label{sec:demon}

From the results presented in the previous section, our proposed MWIN5-RB model does have a better performance. 
Therefore, here we use MWIN5-RB to do the denoising for those data sets with less ADI frames as a further demonstration. 
We try to use this trained MWIN5-RB model on the data sets whose frame numbers are more than 19 and less than 48.
Because the model is already trained, the high quality images made from 
48 frames of data sets are not needed. We only need 20 separated frames and their corresponding 
combined images as the input low quality images.
\begin{figure}
\centering 
\includegraphics[width=0.75\textwidth]{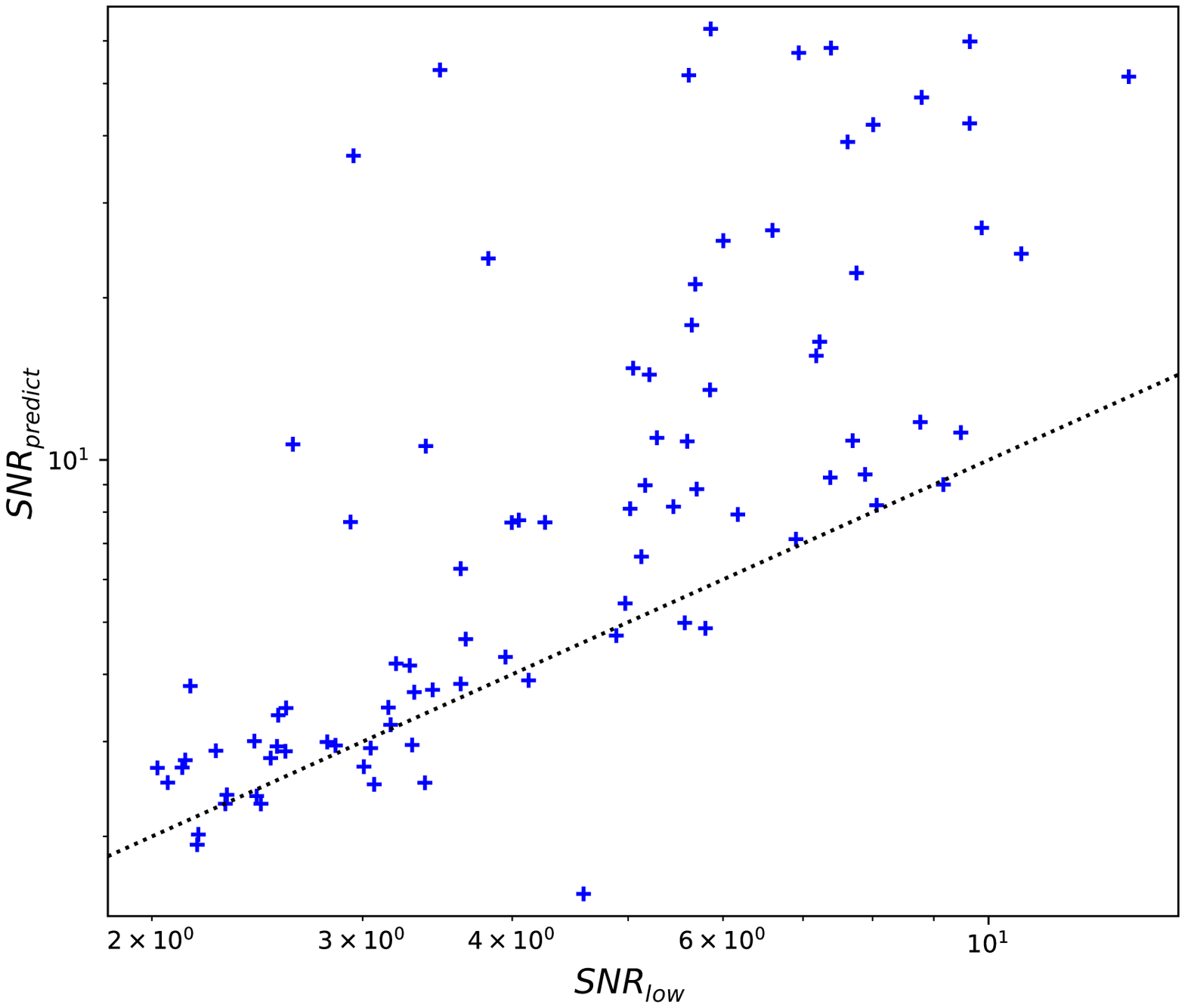}
\caption{The SNR scatter plot of low quality images versus predicted high quality images 
for 88 data sets that have 20-47 frames.
}
\label{fig:risa2047}
\end{figure}
Figure \ref{fig:risa2047} is the scatter plot of the results of 88 data sets.
Among these, the SNRs of predicted high quality images of 72 data sets 
are larger than the SNRs of the corresponding input low quality images.

\begin{figure}
\centering
\includegraphics[width=0.75\textwidth]{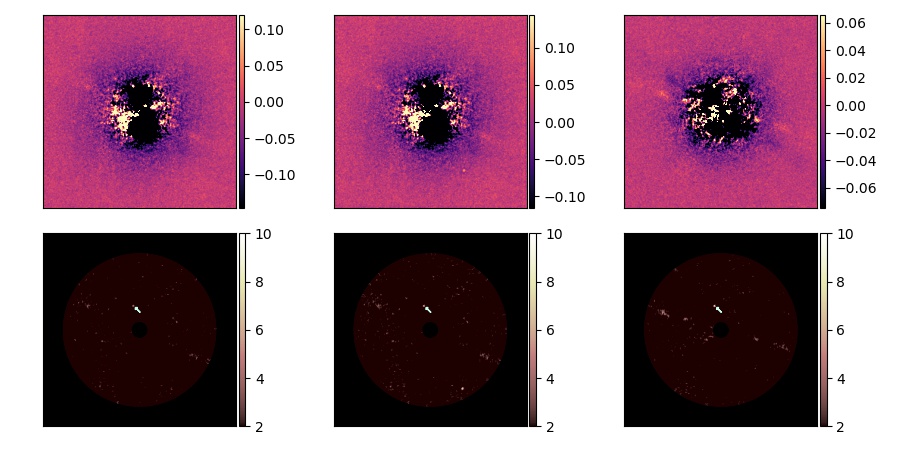}
\caption{The same as Figure \ref{fig:risa1} but for the result of the data set with a known planet HIP 65426b.
From left to right, the SNRs of the upper panels
are 5.878, 7.884, 10.160, respectively.
}
\label{fig:risa_hip65426}
\end{figure}

Furthermore, we also try to use this trained CNN to improve the SNRs of images with a real planet.  
The data set we use for this demonstration is the data with planet HIP 65426b and was taken 
by IRDIS in dual band H2H3 with exposure time 64 seconds on 26th June 2016 \citep{Chauvin.Desidera.2017}.
This data set has 79 frames.
We took 20 ADI frames as the input 
and the results are shown in Figure \ref{fig:risa_hip65426}.
We can see that the SNR of the predicted high quality image is larger than the input low quality image.

\section{Concluding Remarks}
\label{sec:conclu}

The method of direct imaging has made crucial contribution and played an important role in the study of exoplanet.
In order to advance this technique further, building 
larger telescopes is one possibility, and 
improving the methods of image analysis is another. Taking advantages of the development
of machine-learning techniques, here 
we present our study of generating a converter which can increase the SNR of a smaller number of ADI frames of 
the direct-imaging observations of exoplanet search programs. 
In order to perform this investigation, three previously proposed image-denoising 2D-CNN models,
i.e. U-net, DnCNN, WIN5-RB, together with our new MWIN5-RB model are all trained with VLT SPHERE data.
As image converters, 
their denoising results were presented and compared.
It was found that our newly constructed MWIN5-RB model can produce results with higher SNR and increase SNR to be as high as the original high-quality image. 
It was demonstrated 
that this new 2D-CNN model could also increase SNR for the image with a real exoplanet. It was also shown through the SNR maps that our MWIN5-RB does not have the false planet problems of other models seen in Figure \ref{fig:snrmap_compare}.
The only potential limitation is 
that our model would need training data sets in advance. The number
of employed data sets could be similar to what we have used here.
Nevertheless, this can be arranged in a survey program that the first part of 
the survey uses a mode with more ADI frames which could be used as training data sets, 
and the second part uses a different mode 
to search exoplanets by imaging more stars  
with less ADI frames. Therefore, the MWIN5-RB 
model can be trained by the images obtained during the first part of the survey and be employed as a converter to increase the SNR of the images obtained during the second part of the survey.
We therefore conclude that our MWIN5-RB can be useful for those exoplanet search programs which employ the direct imaging technique.

\section*{Acknowledgements}
We are grateful to the anonymous reviewer for 
useful comments.
We thank the European Southern Observatory which makes the VLT data publicly available.
The data employed in this work is
based on the observations collected at the European Southern Observatory under ESO programs 0100.C-0604(A), 0100.D-0399(A), 0101.C-0128(A), 0101.C-0686(A), 0101.C-0753(B), 0102.C-0236(A), 0102.C-0526(A), 095.C-0212(B), 095.C-0298(A), 095.C-0298(B), 095.C-0298(C), 095.C-0298(D), 095.C-0298(I), 095.C-0298(J), 095.C-0346(A), 095.C-0346(B), 095.C-0374(A), 095.C-0426(A), 095.C-0549(A), 095.C-0607(A), 095.C-0838(A), 095.C-0862(A), 096.C-0241(A), 096.C-0241(B), 096.C-0241(C), 096.C-0241(E), 096.C-0241(F), 096.C-0241(G), 096.C-0388(A), 096.C-0414(A), 096.C-0640(A), 096.C-0835(B), 097.C-0060(A), 097.C-0079(A), 097.C-0394(A), 097.C-0523(A), 097.C-0591(A), 097.C-0747(A), 097.C-0750(A), 097.C-0864(A), 097.C-0865(A), 097.C-0865(B), 097.C-0865(C), 097.C-0865(D), 097.C-0865(F), 097.C-0893(A), 097.C-0949(A), 097.C-1006(A), 097.C-1019(A), 097.C-1019(B), 097.C-1042(A), 098.C-0157(A), 098.C-0583(A), 098.C-0686(A), 098.C-0739(C), 098.C-0779(A), 099.C-0177(A), 099.C-0255(A), 099.C-0734(A), 099.C-0878(A), 198.C-0209(A), 198.C-0209(B), 198.C-0209(C), 198.C-0209(D), 198.C-0209(E), 198.C-0209(F), 198.C-0209(G), 198.C-0209(H), 198.C-0209(K), 598.C-0359(F).

\end{document}